\documentclass[12pt]{article}
\usepackage{epsfig}
\usepackage{amsfonts}
\usepackage{amscd}
\usepackage{latexsym}
\usepackage{amsmath,amssymb}
\usepackage{verbatim}
\usepackage{setspace}
\usepackage{color}
\usepackage{cite}

\usepackage[textheight=9in, textwidth=6.5in, letterpaper]{geometry}

\numberwithin{equation}{section}

\def\ip{{\mathcal I}^+}

\def\e{{\epsilon}}
\def\cs{{\cal S}}

 \def\p{\partial}
 \def\bz{{\bar z}}
 
\def\0{{(0)}}
\def\1{{(1)}}
\def\2{{(2)}}
 \def\cL{{\cal L}}
\def\co{{\cal O}}

\def\ci{{\mathcal I}}

\def\<{\langle }
\def\>{\rangle }
\def\bw{{\bar w}}

\def\Q{{\cal Q}}

\newcommand{\bea}{\begin{eqnarray}}
\newcommand{\eea}{\end{eqnarray}}
\newcommand{\be}{\begin{equation}}
\newcommand{\ee}{\end{equation}}
\newcommand{\ba}{\begin{align}}
\newcommand{\ea}{\end{align}}
\def\be{\begin{equation}}
\def\ee{\end{equation}}
\def\beq{\be\begin{array}{c}}
\def\eeq{\end{array}\ee}

\renewcommand{\epsilon}{\varepsilon}

   \makeatletter
  \let\over=\@@over \let\overwithdelims=\@@overwithdelims
  \let\atop=\@@atop \let\atopwithdelims=\@@atopwithdelims
  \let\above=\@@above \let\abovewithdelims=\@@abovewithdelims
\renewcommand\section{\@startsection {section}{1}{\z@}%
                                   {-3.5ex \@plus -1ex \@minus -.2ex}%nn
                                   {2.3ex \@plus.2ex}%
                                   {\normalfont\large\bfseries}}

\renewcommand\subsection{\@startsection{subsection}{2}{\z@}%
                                     {-3.25ex\@plus -1ex \@minus -.2ex}%
                                     {1.5ex \@plus .2ex}%
                                     {\normalfont\bfseries}}

\begin{document}
\begin{titlepage}
\unitlength = 1mm
\ \\
\vskip 1cm
\begin{center}

{ \LARGE {\textsc{Low's Subleading Soft Theorem as a Symmetry of QED }}}

\vspace{0.8cm}
Vyacheslav Lysov,  Sabrina Pasterski and Andrew Strominger

\vspace{1cm}

{\it  Center for the Fundamental Laws of Nature, Harvard University,\\
Cambridge, MA 02138, USA}

\begin{abstract}
  It was shown by F. Low in the 1950s that the subleading terms of soft photon $\cs$-matrix elements obey a universal linear relation. In this paper we give a new interpretation to this old relation, for the case of massless QED, as an infinitesimal symmetry of the $\cs$-matrix.  The symmetry is shown to be locally generated by a vector field on the conformal sphere at null infinity.  Explicit expressions are constructed for the associated charges as integrals over null infinity and shown to generate the symmetry.  These charges are local generalizations of electric and magnetic dipole charges.  
\end{abstract}

\vspace{1.0cm}

\end{center}

\end{titlepage}

\pagestyle{empty}
\pagestyle{plain}

\def\vx{{\vec x}}
\def\p{\partial}
\def\po{$\cal P_O$}

\pagenumbering{arabic}

\tableofcontents
\section{Introduction}

Soft theorems can be reinterpreted as symmetries of the $\cs$-matrix for which the soft particles are Goldstone modes \cite{as,asbms,hms}. A priori, there is no guarantee that the resulting symmetry takes any simple or local form. However, for the case of the soft graviton theorem \cite{steve}, the symmetry turns out to be a diagonal subgroup of the product group of BMS \cite{bms} diffeomorphisms acting on past and future null infinity \cite{hms}.  There is also a subleading soft graviton theorem~\cite{fc},  which is equivalent to a Virasoro symmetry at null infinity~\cite{klps, Adamo:2014yya,Geyer:2014lca}. For the leading soft photon theorem, the resulting symmetry was very recently shown to be the infinite-dimensional subgroup of $U(1)$ gauge transformations which approach the same angle-dependent constant at either end of any light ray crossing Minkowski space \cite{hmps}.  In this paper we consider the subleading soft photon theorem, specializing to massless QED.\footnote{This specialization is made, as in \cite{hmps}, to avoid dealing with singularities in the conformal compactification of past and future timelike infinity in the massive case.}

It has been known since the work of Low~\cite{low54, low}, Burnett-Kroll~\cite{bk} and Gell-Mann-Goldberger \cite{ggm}  that the subleading, as well as the leading, term of soft photon absorption or emission is universal; see equation (3.2) below. In the massless case loop corrections are in general expected~\cite{DelDuca:1990gz,he,bdn}, but we will not consider their effects here.  We re-express the subleading soft relation as a symmetry acting on in- and out-states. However, unlike all the cases mentioned above, the resulting symmetry is $not$ a subgroup of the original gauge symmetry.\footnote{We expect this also to be the case for the subsubleading soft graviton theorem.}  It acts locally on the conformal sphere at $\ci$ where it is parameterized by a vector  field $Y$. However it is bilocal in advanced or retarded time. As already noted in~\cite{he}, the bilocal form is reminiscent of the Yangian appearing in $\mathcal{N}=4$ gauge theories, but we have not found a precise relation. There may also be a connection to the `extra' conserved quantities of Newman and Penrose~\cite{np}.  If $Y$ is one of the global $SL(2,C)$ rotations, the  symmetry implies global magnetic dipole charge conservation.  Having a generic $Y$ is a local generalization of this, in the same sense that supertranslations (superrotations) are local generalizations of global translations (rotations) in the gravity case. 

We wish to stress that, despite the precise formulae presented, the nature 
and significance of the symmetry remains largely mysterious  to us.  It is not a subgroup of the gauge group and, unlike the cases considered in~\cite{as, hmps} does not come under the usual rubric  of asymptotic symmetries.  Moreover, the  infinitesimal symmetry generators do not commute and their commutators give yet more symmetries. We do not know whether or not a finite version of the symmetry transformation exists.  The presence of so many symmetries would ordinarily imply integrability, but it is highly implausible that all abelian theories with massless charges are integrable. Another possibility is that there is no simple extension to massive QED, and loop corrections in the massless case somehow eliminate the symmetries.  Despite all these uncertainties, our formulae seem of interest and are presented here in the hope that further investigations can put them into proper context!

This paper is organized as follows.  In section~\ref{sec:maxwell} we give our conventions, the  mode expansion for the $U(1)$ gauge field and define both the leading and subleading soft operators.  In section~\ref{sec:soft} we review the subleading term in Low's soft photon theorem, and then rewrite it as a symmetry of the $\cs$-matrix.   We construct the associated charges and show that their actions on the fields reproduce the infinitesimal symmetries. The charges are first presented as integrals over all of past or future null infinity, and then, in section~\ref{sec:Qmax}, are shown to reduce to boundary expressions after using the gauge constraints. This is surprising as they are not gauge symmetries! Finally we discuss the connection to dipole charges. 

\section{Preliminaries}
\label{sec:maxwell}
In this section we collect essential formulae and introduce our conventions. For more details see \cite{hmps}.
\subsection{Classical equations}
 Flat Minkowski coordinates $(x^0,x^1,x^2,x^3)$  are given by 
\bea\label{flt}
x^0&=& u+r~~=~v-r,  \cr
  x^1+ix^2&=&{2rz \over 1+z\bz},\cr
  x^3&=& {r(1-z\bz) \over 1+z\bz}.         
  \eea
  where $u$ ($v$) is retarded (advanced) time. 
  In retarded (advanced) coordinates, the metric is 
\be
\label{eq:coord}
ds^2  =-du^2   -2du dr +2r^2 \gamma_{z\bz} dzd\bz =-dv^2   +2dv dr +2r^2 \gamma_{z\bz} dzd\bz , 
\ee
with $\gamma_{z\bz}$ is a round metric on the unit $S^2$. 
In terms of  $\mathcal{F}_{\mu\nu}=\p_\mu\mathcal{A}_\nu-\p_\nu\mathcal{A}_\mu$ and  matter current $j_\nu^M$ the Maxwell equations in retarded coordinates are
\beq
\label{eq:maxwell}
-\gamma_{z\bz} r^2 \p_u \mathcal{F}_{ru} + \p_z \mathcal{F}_{\bz u} + \p_{\bz} \mathcal{F}_{zu} +\p_r (\gamma_{z\bz}r^2\mathcal{F}_{ru} ) = e^2  \gamma_{z\bz}r^2 j_u^M ,\\
\p_z \mathcal{F}_{\bz r} + \p_{\bz}\mathcal{F}_{zr} +\p_r( \gamma_{z\bz} r^2\mathcal{F}_{ru}) =e^2 \gamma_{z\bz} r^2 j_r^M,\\
r^2 \p_r(\mathcal{F}_{rz} -\mathcal{F}_{uz}) -r^2 \p_u \mathcal{F}_{rz} +\p_z (\gamma^{z\bz} \mathcal{F}_{\bz z}) = e^2 r^2 j_z^M.
\eeq
A similar expression applies to advanced coordinates. 

\subsection{Mode expansions}

The  mode expansion for the outgoing free Maxwell field is 
\begin{equation}
\mathcal{A}^{out}_{\mu}(x) =e \sum\limits_{\alpha=\pm} \int \frac{ d^3q}{(2\pi)^3} \frac{1}{2 \omega_q} \left( \e^{\alpha*}_{\mu} ({ \vec q})a^{out}_\alpha ({\vec q}) e^{i q \cdot x} + \e^\alpha_{\mu}({\vec q}) a^{out}_\alpha ({\vec q})^\dagger   e^{- i q \cdot x} \right),
\end{equation}
where $q^0 = \omega_q = | {\vec q}|$, $\alpha=\pm$ are the two helicities and
\be\label{rrd}
[a^{out}_\alpha ({\vec q}), a^{out}_\beta ({\vec{q'}})^\dagger ]=  2\omega_q\delta_{\alpha\beta} (2\pi)^3 \delta^3 \left( {\vec q} - {\vec q}' \right).
\ee 
Outgoing photons with momentum $q$  and helicity $\alpha$ correspond to final-state insertions of $a^{out}_\alpha ({\vec q})$.  They arrive at a point $w$ on the conformal sphere at $\ci^+$. It is convenient to parametrize the photon four-momentum  
by  $(\omega_q,w,\bw)$ \begin{equation}\label{gravmom}
q^\mu = \frac{\omega_q}{1 + w {\bar w}} \left( 1 + w {\bar w} , w + {\bar w} ,  i \left( \bar{w} -  w\right), 1 - w {\bar w}  \right),
\end{equation}
with  polarization tensors 
\beq \label{gg}
{ \e}^{+\mu}( {\vec q} ) = \frac{1}{\sqrt{2}} \left( {\bar w}, 1, - i, - {\bar w} \right),  \\
{\e}^{-\mu}({\vec q} ) = \frac{1}{\sqrt{2}} \left( w , 1,   i, - w  \right).
\eeq
These obey  $\e^{\pm\mu}q_\mu=0$ and 
\be \e_{\bz}^+ \left(\vec{q} \right) = \p_{\bz} x^\mu\e^+_\mu \left( \vec{q} \right)= \frac{ \sqrt{2} r \left( 1 + z\bar{w} \right)}{ \left( 1 + z {\bar z} \right)^2 }   ,~~~~\e_{\bz}^- \left({\vec q} \right) = \p_{\bz} x^\mu\e^-_\mu \left( \vec{q} \right)  = \frac{ \sqrt{2} r { z} \left( { w} - { z} \right) }{ \left( 1 + z {\bar z} \right)^2 }.\ee
We define the boundary field on $\ci^+$ by
 \begin{equation}
A_{\bz}(u,z,\bz) =  \lim_{r\to\infty}  \mathcal{A}^{out}_{\bz}(u,r,z,\bz)=\lim_{r\to\infty}  \p_\bz x^\mu \mathcal{A}^{out}_{\mu}(u,r,z,\bz).
\end{equation}
This is related to the plane wave modes by 
\be
A_{\bz} = e \lim\limits_{r\to\infty} \p_\bz x^\mu   \sum\limits_{\alpha=\pm} \int \frac{ d^3q}{(2\pi)^3} \frac{1}{2 \omega_q} \left( \e^{\alpha*}_{\mu} ({ \vec q})a^{out}_\alpha ({\vec q}) e^{- i \omega_q u - i \omega_q r \left( 1 - \cos\theta \right) } + h.c. \right)
\ee
where $\theta$  is the angle between  between the $\vec{x}$ and $\vec{p}$.
At large $r$ the leading saddle point approximation near $\theta=0$ gives
\be
\label{eq:mode}
A_{\bz} =  - \frac{i e\hat{\e}_{\bz}^+}{8\pi^2}  \int\limits^\infty_0   d\omega_q  ( a^{out}_- (\omega_q \hat{x}) e^{- i \omega_q u} -
 a^{out}_+ (\omega_q \hat{x})^\dagger e^{ i \omega_q u} ).
\ee
Here, $\hat{x}$ is parameterized by $(z,\bar{z})$
\be
\hat{x}\equiv \frac{\vec{x}}{r} = \frac{1}{1+z\bz} (z+\bz, i (\bz-z), 1-z\bz)
\ee
 and 
\be
\hat{\e}^{+ }_{\bz} = \frac{\p_\bz x^\mu }{r} \e_{\mu}^{+}  = \frac{\sqrt{2}}{1+z\bz}.
\ee
One may also check that in the gauge \eqref{gg}, $A_u=\lim\limits_{r\to \infty} \p_u x^\mu \mathcal{A}^{out}_\mu$ vanishes on $\ci^+$ and hence
$F_{u\bz}(u,z,\bz)= \p_u A_{\bz}(u,z,\bz)$.
Using~(\ref{eq:mode}), a similar mode expansion for $A_{z}$, and the commutation relations~(\ref{rrd})  the $\ci^+$ commutator is \be\label{AAcomm}
\big[F_{u\bz}(u,z,\bz), F_{u'w}(u',w,\bw)\big]   = \frac{ie^2}{2} \delta^2 (z-w)\p_u\delta (u-u').
\ee
Similarly, defining the field $A^-_\bz $ on $\mathcal{I}^-$ by
\be
\label{eq:mode2}
A^-_{\bz} =  - \frac{i e\hat{\e}_{\bz}^+}{8\pi^2}  \int\limits^\infty_0   d\omega_q  ( a^{in}_- (\omega_q \hat{x}) e^{- i \omega_q v} -
 a^{in}_+ (\omega_q \hat{x})^\dagger e^{ i \omega_q v} ),
\ee
gives
\be\label{Gcomm}
\big[G_{v\bz}(v,z,\bz), G_{v'w}(v',w,\bw)\big]   = \frac{ie^2}{2} \delta^2 (z-w)\p_v\delta (v-v'),
\ee
where $G_{vz} =\p_v A^-_z$.

\subsection{Soft photon operators}
We would now like to construct the operators corresponding to soft photon insertions on $\mathcal{I}^+$ and $\mathcal{I}^-$.  To examine the soft limit of the above mode expansions, we define
\be
\begin{array}{ll}
F^{\omega}_{u\bz}&\equiv \int du e^{i\omega u} \p_u A_{\bz} \\
&= - \frac{e}{4\pi}\hat{\e}_{\bz}^+ \int\limits^\infty_0   \omega_q d\omega_q  [ a^{out}_- (\omega_q \hat{x}) \delta(\omega-\omega_q) + a^{out}_+ (\omega_q \hat{x})^\dagger \delta(\omega+\omega_q) ].
\end{array}
\ee
For $\omega>0$ only the first delta-function contributes, while for $\omega<0$ only the second:
\begin{equation}
\begin{array}{lll}
F_{u\bz}^\omega&=& - \frac{ e  }{4\pi }\hat{ \e}_{\bz}^{+}\omega a^{out}_- (\omega {\hat x }) ,  \\
F_{u\bz}^{-\omega}&=& - \frac{ e  }{4\pi } \hat{ \e}_{\bz}^{+}\omega a^{out}_+ (\omega {\hat x })^\dagger, 
\end{array}
\end{equation}
with $\omega>0$ in both cases.
Similarly on  $\mathcal{I}^-$
\begin{equation}
\begin{array}{lll}
G_{v\bz}^\omega&=& - \frac{ e }{4\pi }  \hat{ \e}_{\bz}^{+} \omega a^{in}_- (\omega {\hat x }),   \\
G_{v\bz}^{-\omega}&=& - \frac{  e }{4\pi } \hat{ \e}_{\bz}^{+}\omega a^{in}_+ (\omega {\hat x })^\dagger. 
\end{array}
\end{equation}
The zero mode of $F_{u\bz}$ is defined as
\be
\label{eq:zF}
\begin{array}{ll}
F^0_{u\bz}&\equiv \frac12 \lim\limits_{\omega\to 0}(F^\omega_{u\bz}+F^{-\omega}_{u\bz} ) \\
&=-\frac{e}{8\pi} \hat{\e}_{\bz}^{+}\lim\limits_{\omega\to  0}[\omega a^{out}_-(\omega\hat{x})+ \omega a^{out}_+(\omega\hat{x})^\dagger],\\
\end{array}
\ee while on $\mathcal{I}^-$
\be
\begin{array}{ll}
G^0_{v\bz}&\equiv \frac12 \lim\limits_{\omega\to 0}(G^\omega_{v\bz}+G^{-\omega}_{v\bz} )\\
&=-\frac{e }{8\pi}\hat{\e}_{\bz}^{+} \lim\limits_{\omega\to  0}[\omega a^{in}_-(\omega\hat{x})+ \omega a^{in}_+(\omega\hat{x})^\dagger].
\end{array}
\ee
As in~\cite{klps}, it is useful to define operators which create subleading soft photons, insertions of which automatically have the soft pole projected out.  These are given on $\ci^+$ by
\be
 \begin{array}{ll}
F^{(1)}_{u\bz} &\equiv \int du~u \p_u A_{\bz} \\
&=  - \lim\limits_{\omega \to 0}\frac{i}{2}(\p_\omega F^\omega_{u\bz}+\p_{-\omega}  F^{-\omega}_{u\bz})\\
&=\frac{ie}{8\pi} \hat{\e}_{\bz}^{+} \lim\limits_{\omega\rightarrow0}(1+\omega\partial_\omega)[ a^{out}_-(\omega\hat{x})-a^{out}_+(\omega\hat{x})^\dagger],
\end{array}
\ee
while at $\mathcal{I}^-$
 \be
 \begin{array}{ll}
G^{(1)}_{v\bz} &\equiv \int dv~v \p_v A^-_{\bz} \\
&=  - \lim\limits_{\omega \to 0}\frac{i}{2}(\p_\omega G^\omega_{v\bz}+\p_{-\omega}  G^{-\omega}_{v\bz})\\
&=\frac{ie}{8\pi} \hat{\e}_{\bz}^{+} \lim\limits_{\omega\rightarrow0}(1+\omega\partial_\omega)[ a^{in}_-(\omega\hat{x})-a^{in}_+(\omega\hat{x})^\dagger].
\end{array}
\ee

\section{Soft theorem $\to$ symmetry}
\label{sec:soft} 
In this section we rewrite the subleading soft theorem as an asymptotic symmetry acting on in- and out-states. 
Let us denote a state with $n$ massless hard particles of energies $E_k$, charges $eQ_k$ and momenta 
\be \label{mom}
p_k^\mu = \frac{E_k}{1 + z_k {\bar z_k}} \left( 1 + z_k {\bar z_k} , z_k + {\bar z_k} ,  i \left( \bar{z}_k -  z_k\right), 1 - z_k {\bar z_k}  \right),
\end{equation}
by $|z_1,...\rangle$, and hard  $\cs$-matrix elements by $\langle z_{n+1},...|  \cs |z_1,...\rangle$.
The Low-Burnett-Kroll-Goldberger-Gell-Mann soft theorem~\cite{low54,low,bk,ggm, ca, ad,zvi} then states that if we add to the out-state a positive helicity photon with energy 
$\omega \to 0$, the first two terms in the soft expansion are 
\be
\langle z_{n+1},...| a^{out}_-(\vec q) \cs |z_1,...\rangle   = (J^{(0)-}   + J^{(1)-}  ) \langle z_{n+1},...|  \cs  |z_1,...\rangle + \co(\omega).
\ee
Here
\be \label{zz}
J^{(0)-} = e\sum\limits_k  Q_k\frac{p_k\cdot \e^-}{p_k\cdot q}\sim \co(\omega^{-1}),\;\;\;J^{(1)-}=-ie\sum\limits_k  Q_k \frac{ q_\mu\e_\nu^-J_k^{\mu\nu}} { p_k\cdot q}\sim \co(\omega^{0}), \ee
with $J_{k\mu\nu} $ the total angular momentum operator of the $k^{th}$ particle.  In \cite{hmps} it was shown that the leading $J^{(0)}$ term implies a symmetry under large gauge transformations which approach an arbitrary angle dependent gauge transformation at null infinity. Here we wish to understand the subleading $J^{(1)}$ term. For this purpose it is convenient to eliminate the $J^{(0)-}$ contribution using the projection operator $(1+\omega \p_\omega)$
\be
\lim_{\omega\to 0} (1+\omega\p_\omega)\langle z_{n+1},...| a^{out}_-(\vec q) \cs |z_1,...\rangle   = J^{(1)-}  \langle z_{n+1},...|  \cs  |z_1,...\rangle .
\ee
From~(\ref{eq:zF}) one than has 
\be\label{ss}
\begin{array}{ll}
e\hat{\e}^+_{\bz}J^{(1)-}  \langle z_{n+1},...|  \cs  |z_1,...\rangle&=
e\hat{\e}^+_{\bz}\lim\limits_{\omega\to 0} (1+\omega\p_\omega)\langle z_{n+1},...| a^{out}_-(\vec q) \cs |z_1,...\rangle  \\
&=-8\pi i \langle z_{n+1},...| F^{(1)}_{u\bz} \cs |z_1,...\rangle . 
\end{array}
\ee
For the special case of a scalar field with $J_{k\mu\nu}=-i\left(p_{k\mu}{\p \over \p p_k^\nu}- p_{k\nu}{\p \over \p p_k^\mu}\right) $, rewriting ($p_k^\mu, q^\mu$) in terms of $(E_k, z_k, \bar{z}_k)$ in \eqref{zz} gives 
for the right hand side of \eqref{ss}    \be
 J^{(1)-}=
  -e\sum_k  \frac{  Q_k}{\sqrt{2}   ( \bz_k - \bar{z} ) } \left[(1+z\bz_k)\p_{E_k}+ E_k^{-1} (z-z_k)(1+z_k\bar{z}_k) \p_{z_k} \right].
 \ee
 This is nonlocal on the conformal sphere. However 
 acting with two covariant derivatives gives the local expression
 \be
  D_z^2 (\hat{\e}_\bz^+J^{(1)-}) = 2\pi e\sum_k  Q_k  \left(   D_z \delta^2(z-z_k)\p_{E_k}
+E_k^{-1} \delta^2 (z-z_k) \p_{z_k}\right).
 \ee
Acting with  $D_z^2$  on  both sides of the soft theorem and integrating  the result against an arbitrary vector field $Y^z$ gives\footnote{\label{bound}Various conditions at the boundaries of $\ci$ may lead one to impose constraints such as $D_\bz D_z^2Y^z=0$.}
 \beq
\int d^2z \;D_z^2 Y^ze\hat{\e}^+_{\bz}\lim\limits_{\omega\to 0} (1+\omega\p_\omega)\langle z_{n+1},...| a^{out}_-(\vec q) \cs |z_1,...\rangle~~~~~~~~~~~~~~~~~~~~~~~~\\  
~~~~~~~~~~~~~~~~~~~~~~=
-2\pi e^2 \sum\limits_k  Q_k\left( D_zY^z (z_k)\p_{E_k}-E_k^{-1} Y^{z}(z_k)\p_{z_k}\right) \langle z_{n+1},...|  \cs  |z_1,...\rangle.
 \eeq
 For spinning  fields we need to replace  $Y^z\p_{z}$ by the  Lie derivative $\cL_Y$. For a  hermitian action we should include $\hat \e_z^-$ and $Y^\bz$ but we suppress this for notational brevity. 
Similarly for the insertion of an incoming soft photon 
  \beq
-\int d^2z \;D_z^2 Y^{z}e\hat{\e}^+_{\bz}\lim\limits_{\omega\to 0} (1+\omega\p_\omega)\langle z_{n+1},...| \cs a^{in}_+(\vec q)^\dagger |z_1,...\rangle~~~~~~~~~~~~~~~~~~~~~~~~\\  
~~~~~~~~~~~~~~~~~~~~~~=
-2\pi e^2 \sum\limits_k  Q_k\left( D_zY^{z} (z_k)\p_{E_k}-E_k^{-1} Y^{z}(z_k)\p_{z_k}\right) \langle z_{n+1},...|  \cs  |z_1,...\rangle.
 \eeq
Let us  define soft photon operators
  \be
\Q^+_S = -\frac{2}{e^2} \int d^2z du\;  u\p_uA_{\bz} D^2_z Y^z,  
\ee
  \be
\Q_S^- =  \frac{2}{e^2} \int  d^2z dv\;   v\p_vA^-_{\bz} D^2_z Y^{z}.  
\ee
Hard particle symmetry operators  $\Q^\pm_H$ are defined by their action 
 \be
 \label{eq:qh}
 \begin{array}{ll}
\langle E,z|   \Q^+_H&= -i  Q \left( D_zY^z \p_{E}-E^{-1} Y^{z}\p_{z}\right)\langle E,z|,\\
 \end{array}
 \ee
 \be
 \label{eq:qha}
 \begin{array}{ll}
 \Q^-_H| E,z\rangle &= i Q  \left( D_zY^{z} \p_{E}-E^{-1} Y^{z}\p_{z}\right)| E,z\rangle.\\
 \end{array}
 \ee
Finally we write 
\be  \Q^\pm=\Q^\pm_S+\Q^\pm_H. \ee
 Then the subleading soft theorem for massless QED takes the form 
 \be
  \langle z_{n+1},...| \Q^+\mathcal{S} - \mathcal{S} \Q^-   |z_1,...\rangle =0.
 \ee
 This expresses the subleading term in Low's theorem as  an infinitesimal symmetry of the massless QED $\cs$-matrix.

   \section{Charges}
   \label{sec:Qmax}

  In this section we express the operators $\Q^\pm$, for the case of scalar charged fields,  as integrals over local fields on $\ci^\pm$. The fact that this is possible is perhaps  surprising as the factor of $E^{-1}$ in \eqref{eq:qh} suggests time nonlocality. 
   
  A massless scalar field has an expansion near $\ip$
 \be
 \Phi(u,r,z,\bz) = \frac{\phi(u,z,\bz)}{r} + \sum_{n=0}^\infty\frac{\phi^{n}(u,z,\bz)}{r^{n+2}}.
 \ee
 The commutation relation for the boundary field at $\ip$ is 
 \be \label{cc}
 [\phi (u,z,\bz), \bar{\phi}(u',w,\bw)] = -\frac{i\gamma^{z\bz}}{4} \Theta (u-u') \delta^2(z-w),
 \ee
 where $\Theta(x)$ is  the sign function.
 The boundary charge  current is 
\be
J_\mu^M = i Q \lim_{ r \to \infty} r^2 (\bar{\Phi} \p_\mu \Phi  - \Phi \p_\mu \bar{\Phi}) = iQ(\bar{\phi} \p_\mu \phi  - \phi \p_\mu \bar{\phi}). 
\ee
Expressing $ \Q^+_H$ in terms of current operators gives   \be
 \Q^+_H  = \int_{\ip}d^2zdu( u D_zY_\bz J^M_u +Y_\bz J^M_z).
 \ee
 Using \eqref{cc} as well as 
 \be
\frac{i}{\pi}\int \frac{e^{-iEu}}{E+i\epsilon^+} dE = 1+\Theta (u),
 \ee
 one finds the desired action on the Fourier transform 
 $ \phi_E = \int du\; e^{iEu} \phi $ of $\phi$
 \be \label{cc2}
 [\mathcal{Q}^+_H, {\phi}_E(z,\bz)] = iQ  \left( D_zY^{z} \p_{E}-E^{-1} Y^{z}\p_{z}\right) {\phi}_E(z,\bz).
 \ee
 Similarly on $\mathcal{I}^-$
 \be
 \Q^-_H  = -\int_{\mathcal{I}^-}d^2zdv(v D_zY_\bz J^M_v +Y_\bz J^M_z)
 \ee
 generates the hard action \eqref{eq:qha} on incoming massless scalars. 
 It is likely possible to generalize the construction to spinning fields but we have not worked out the details. 

Using the constraint equations~(\ref{eq:maxwell}), one can eliminate the matter charge currents and express the combined hard and soft charges as a boundary term.  On $\ci^+$
\be
\Q^+=\lim\limits_{r\rightarrow\infty}\frac{1}{e^2}\int_{\ci^+}dud^2z\p_u\big(uD_zY^z(r^2\mathcal{F}_{ur}\gamma_{z\bz}+\mathcal{F}_{z\bz})+2r^2Y_\bz\mathcal{F}_{zr}\big).
\ee
For the field configurations that revert to vacuum at  $\mathcal{I}^+_+$ this reduces to the $S^2$ integral 
\be \Q^+=-\lim\limits_{r\rightarrow\infty}\frac{1}{e^2}\int_{\ci^+_-}d^2z\big(uD_zY^z(r^2\mathcal{F}_{ur}\gamma_{z\bz}+\mathcal{F}_{z\bz})+2r^2Y_\bz\mathcal{F}_{zr}\big).\ee
Similarly on $\ci^-$ 
\be
\Q^-=\lim\limits_{r\rightarrow\infty}\frac{1}{e^2}\int_{\ci^-_+}d^2z\big(vD_zY^z(r^2\mathcal{F}^-_{vr}\gamma_{z\bz}-\mathcal{F}^-_{z\bz})+2r^2Y_\bz\mathcal{F}^-_{zr}\big).
\ee
It is interesting to compare these to the expressions for the electric and magnetic  charges $Q$ and $\tilde Q$ and  the dipole moments $\vec{\wp}$ and  $\vec{\mu}$:  
\be
e^2Q+2\pi i\tilde Q = \lim\limits_{r\rightarrow\infty}\int d^2z\; ( r^2\mathcal{F}_{ru}\gamma_{z\bz}+\mathcal{F}_{z\bz})\\
\ee
\be
-e^2\vec{\wp} + 2\pi i\vec{\mu} = \lim\limits_{r\rightarrow\infty}3\int d^2z\; r^2 \mathcal{F}_{zr}\p_{\bz} \hat{x}.
\ee

We see that if we take $Y$ to be a global $SL(2,C)$ rotation and use the boundary condition $\mathcal{F}_{z\bz}$=0 from~\cite{hmps},\footnote{For such rotations, $Y$ is real and hence entails nonzero $Y^\bz$ which we have been suppressing.  We note that the particular restriction on $Y$ mentioned in footnote~\ref{bound} would eliminate these rotations.} $\Q^\pm$ are nothing but the total magnetic dipole charge. This is `conserved' in the sense that, given that the system begins and ends in the vacuum, the total incoming dipole charge must equal the total outgoing dipole charge. More generally, $\Q^\pm$ are local generalizations of dipole charge in the same sense that supertranslations (superrotations) are local generalizations of global translations (rotations). Hence the conservation law that implies Low's subleading soft theorem may be heuristically thought of as  the equality of total incoming and total outgoing  dipole charge flux at every fixed  angle.

\section*{Acknowledgements}
We are  grateful  to F. Cachazo, T. He, P. Mitra, and M. Schwartz for useful conversations. This work was supported in part by DOE grant DE-FG02-91ER40654 and the Fundamental Laws Initiative at Harvard.

\end{document}